\documentclass[a4paper]{jpconf}
\usepackage{graphicx}

\def\beq{\begin{equation}}
\def\eeq{\end{equation}}
\def\beqa{\begin{eqnarray}} 
\def\eeqa{\end{eqnarray}}
\def\laq{\raise 0.4 ex \hbox{$<$}\kern -0.8 em\lower 0.62 ex\hbox{$\sim$}}
\def\gaq{\raise 0.4 ex \hbox{$>$}\kern -0.7 em\lower 0.62 ex\hbox{$\sim$}}

\begin{document}
\title{Noncommutative effects in astrophysical objects: a survey}

\author{Orfeu Bertolami$^{1,2}$ and Carlos A D Zarro$^{1,2}$}

\address{$^1$ Departamento de F\'\i sica, Instituto Superior T\'ecnico, \\
Av. Rovisco Pais, 1049-001 Lisboa, Portugal}
\address{$^2$ Instituto de Plasmas e Fus\~{a}o Nuclear, Instituto Superior T\'ecnico,
Av. Rovisco Pais, 1049-001 Lisboa, Portugal}

\ead{orfeu@cosmos.ist.utl.pt, carlos.zarro@ist.utl.pt}

\begin{abstract}
The main implications of noncommutativity over astrophysical objects are examined. Noncommutativity is introduced through a deformed dispersion relation $E^{2}=p^{2}c^{2}(1+\lambda E)^{2} + m^{2}c^{4}$ and the relevant thermodynamical quantities are calculated using the grand canonical ensemble formalism. These results are applied to simple physical models describing main-sequence stars, white-dwarfs and neutron stars. The stability of main-sequence stars and white dwarfs is discussed.
\end{abstract}

\section{Introduction}

Noncommutativity is believed to be an important feature of space-time at quantum gravity scales \cite{Connes:1996gi,Madore:1999bi}. Interestingly, it also appears in the context of string theory \cite{Witten:1985cc,Seiberg:1999vs}.

Noncommutative quantum field theory (NCQFT) models can be implemented by substituting the normal product between fields by a noncommutative (NC) one, the so-called Moyal product \cite{Szabo:2001kg,Douglas:2001ba}. This procedure introduces a minimum length scale, but does not solve, as one could expect, the renormalization problem of the quantum field theory. One also encounters additional difficulties on issues such as causality and unitarity. Related questions associated with noncommutativity involve the violation of translational invariance \cite{Bertolami:2003nm}, scalar fields and their stability in curved spaces \cite{Lizzi:2002ib,Bertolami:2002eq, Bertolami:2008zv}. NCQFT can also be implemented by generalizing the Heisenberg-Weyl algebra of noncommutative quantum mechanics \cite{Zhang:2004yu,Bertolami:2005jw,Acatrinei:2003id,Bertolami:2005ud,Bastos:2006kj,Bastos:2006ps,Gamboa:2001fg,Snyder:1947} to field algebra \cite{Gamboa:2001fg}. This approach can be used, for instance, to investigate the inflationary period \cite{Barosi:2008gx}. Another way to introduce noncommutativity in field theory is through a deformed dispersion relation in a commutative space-time rather than considering a noncommutative space-time over which the fields are defined \cite{Alexander:2001ck}. Since there is no consistent theory of noncommutative gravity \cite{AlvarezGaume:2006bn,Meyer:2005as} this approach is justified. This is inspired by quantum groups methods \cite{AmelinoCamelia:1999pm}, by a possible breaking of Lorentz symmetry at high-energies \cite{Bertolami:1999da,Bertolami:2003yi,Bertolami:2003tw} and the modifications of special relativity that take into account the possible existence of a minimum length \cite{Magueijo:2001cr,AmelinoCamelia:2000mn}.

One mentions here some possible ways on how noncommutativity can lead to a deformed dispersion relation. The first approach is related to quantum groups methods \cite{AmelinoCamelia:1999pm}. Assuming the following commutation relations 

\beq\label{eq:1}
[x^{i},t]=i\lambda x^{i}\;\;\;\;\;\;\;\;\;\;\;\;\;\; [x^{i},x^{j}]=0,
\eeq

\noindent where $i,j=1,2,3$ and $\lambda$ is constant, it can be shown that the associated relationship between dynamical quantities is given by \cite{AmelinoCamelia:1999pm}

\beq\label{eq:2}
\lambda^{-2}(e^{\lambda E}+e^{-\lambda E}-2) - p^{2}c^{2}e^{-\lambda E}=m^{2}c^{4},
\eeq

\noindent which is a deformed dispersion relation. 

A relationship between noncommutativity and a deformed dispersion relation also arises in the context of special relativity proposal with an invariant length \cite{Magueijo:2001cr}. In this instance, one modifies the boost generator and from the invariant quantity 

\beq\label{eq:smolinmagueijo}
||p||^{2}=m^{2}c^{4}=\frac{p^{\mu}p_{\mu}}{(1-L_{P}E)^{2}}, 
\eeq

\noindent where $L_{P}$ is the invariant length,  a deformed dispersion relation follows. The thermodynamics of an ideal fluid following the deformed dispersion relation Eq. (\ref{eq:smolinmagueijo}) was investigated in Ref. \cite{Das:2009qb}.

From the previous discussion, it is natural to expect that NC arises at high energies, and hence in the early universe \cite{Alexander:2001ck,Alexander:2001dr}. However, in this contribution one discusses the results of Ref. \cite{Bertolami:2009wa} where NC low-energy effects on  astrophysical objects are investigated. Also here the noncommutativity is introduced via a deformed dispersion relation. Using the grand canonical ensemble, one finds the leading noncommutative correction to  particle number, energy densities and pressure. This method is applied to radiation, non-relativistic ideal gas and degenerate fermion gas and then used to describe the main features of main-sequence stars, white dwarfs and neutron stars.


\section{Deformed Dispersion Relation}\label{sec:defdisprel}

One can define a simplified version of Eq. (\ref{eq:2}) as in Ref. \cite{Alexander:2001ck}

\beq \label{eq:defdisprel}
E^{2}=p^{2}c^{2}(1+\lambda E)^{2} + m^{2}c^{4}.
\eeq

\noindent Indeed, performing $\lambda	\rightarrow 2\lambda$ in Eq. (\ref{eq:2}), Eq. (\ref{eq:defdisprel}) matches Eq. (\ref{eq:2}) up to the first order in $\lambda$ \cite{Alexander:2001ck}. Solving for $E$ one gets 

\beq \label{eq:completeE}
E=\frac{\lambda p^{2}c^{2} + \sqrt{p^{2}c^{2} + m^{2}c^{4}(1-\lambda^{2}p^{2}c^{2})}}{1-\lambda^{2}p^{2}c^{2}}.
\eeq

\begin{figure}
\centering
\includegraphics[width=7cm]{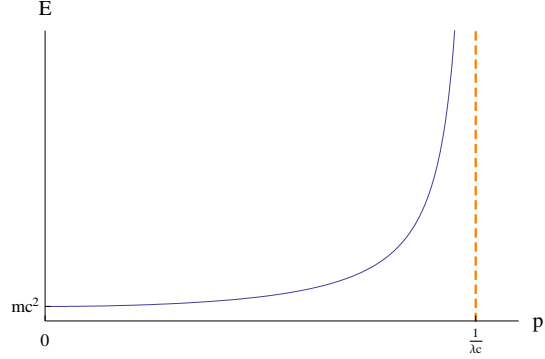}
\caption{Deformed dispersion relation for the particle branch of Eq. (\ref{eq:completeE}).}
\label{fig:defdisprel}
\end{figure}

The general behaviour of Eq. (\ref{eq:completeE}) is depicted in Fig. \ref{fig:defdisprel}. One considers the particle branch, where all energies can be attained although there is a maximum momentum given by $p_{max}=1/\lambda c$. Thus the parameter $\lambda$ is associated with the maximum momentum \cite{Alexander:2001dr}. Since one aims to investigate energy scales encountered in astrophysical objects and  $\lambda$ is presumably related to the inverse of the quantum gravity energy scale, it suffices to keep only the first correction in $\lambda$. Therefore, at first order in $\lambda$, Eq. (\ref{eq:completeE}) reads

\beq\label{eq:folambda}
E = \lambda p^2c^{2} + \sqrt{p^{2}c^{2}+m^{2}c^{4}}.
\eeq 

\noindent Notice that the usual relativistic dispersion relation is recovered for $\lambda \rightarrow 0$.


\section{Deformed Statistical Mechanics}\label{sec:defstatmech}

The foundations of statistical mechanics are not modified for more general dispersion relations \cite{Alexander:2001ck,Landau:1980,Padmanabhan:2000}. One considers the grand canonical ensemble, since it is the suitable formalism to describe radiation, ideal gas and degenerate fermion gas.  Consider a system with $N$ particles. Each state is labeled by $j$ ($j=1,2,\cdots$) and corresponds to $n_{j}$ particles with energy $E_{j}$. One defines the fugacity as $e^{\beta \mu}$, where $\mu$ is the chemical potential, $\beta=(k_{B}T)^{-1}$, $k_{B}$ is the Boltzmann constant and $T$ the absolute temperature. The grand canonical partition function is defined as 

\beq\label{eq:gcpf}
\mathcal{Z}=\sum_{n_{j}}\prod_{j}\left[z e^{-\beta E_{j}} \right]^{n_{j}}=\sum_{n_{j}}\prod_{j}\left[e^{\beta(\mu- E_{j})} \right]^{n_{j}}.
\eeq

\noindent Through the grand canonical potential, $\Phi$, one obtains the thermodynamics \cite{Landau:1980}

\beq\label{eq:gcpot}
\Phi=-PV=-\frac{1}{\beta}\ln\mathcal{Z}=-\frac{1}{a\beta}\sum_{j}\ln(1+aze^{-\beta E_{j}})
\eeq

\noindent where $P$ is the pressure, $V$ the volume, $a=1$ for fermions and $a=-1$ for bosons. Using Eq. (\ref{eq:gcpot}), the pressure is obtained in  the large-volume limit $\sum_{E} \rightarrow \int{\frac{d^{3}\vec{x}d^{3}\vec{p}}{(2\pi\hbar)^{3}}}$,

\beq\label{eq:gcpotc}
\Phi=-PV=-\frac{\gamma}{a\beta}\int{\frac{d^{3}\vec{x}d^{3}\vec{p}}{(2\pi\hbar)^{3}}}\ln(1+aze^{-\beta E(\vec{x},\vec{p})}),
\eeq

\noindent where $E=E(\vec{x},\vec{p})$ and $\gamma$ is the multiplicity of states due to spin. At this point one has to introduce an specific relationship for $E=E(\vec{x},\vec{p})$ and one chooses Eq. (\ref{eq:completeE}) for this function. Since the deformed dispersion relation Eq. (\ref{eq:completeE}) depends only on the absolute value of the momentum, one can compute the configuration variable integral in Eq. (\ref{eq:gcpotc}), which yields the volume $V$. The pressure then reads

\beq\label{eq:pressure}
P=\frac{\gamma}{2\pi^{2}\hbar^{3}}\int_{0}^{\frac{1}{\lambda c}}dp \left(\frac{p^{2}}{z^{-1}e^{\beta E} + a} \right)\left(\frac{p}{3}\frac{dE}{dp}\right).
\eeq    

Notice that the upper limit of this integral is $p_{max}=(\lambda c)^{-1}$. The average number of particles and the average energy are given respectively by

\beq\label{eq:anaep}
\langle N \rangle=-\left(\frac{\partial \Phi}{\partial \mu}\right)_{\beta}=\sum_{E}n(E), ~~~~~~~~\langle E \rangle = \sum_{E} E\; n(E),
\eeq

\noindent where $n(E)=\left(z^{-1}e^{\beta E} + a \right)^{-1}$ is the occupation number of particles with energy $E$. Considering the large-volume limit and integrating, one finds the particle number and energy densities

\beq\label{eq:numberdensity}
\frac{N}{V}=\frac{\gamma}{2\pi^{2}\hbar^{3}}\int_{0}^{\frac{1}{\lambda c}}dp \left(\frac{p^{2}}{z^{-1}e^{\beta E} + a} \right),
\eeq    

\beq\label{eq:energydensity}
u=\frac{E}{V}=\frac{\langle E \rangle}{V}=\frac{\gamma}{2\pi^{2}\hbar^{3}}\int_{0}^{\frac{1}{\lambda c}}dp\;p^{2} \left(\frac{E}{z^{-1}e^{\beta E} + a} \right).
\eeq

\subsection{Deformed Radiation}
A gas of photons is used to describe radiation. For photons, the deformed dispersion relation Eq. (\ref{eq:defdisprel}) reads

\beq\label{eq:Erad}
E=pc(1+\lambda E).
\eeq

\noindent To obtain the energy density one substitutes Eq. (\ref{eq:Erad}) into Eq. (\ref{eq:energydensity}) with $a=-1$, $\mu=0$ and $\gamma=2$, that is:

\beq\label{eq:uintdef}
u=\frac{1}{\pi^{2}\hbar^{3}c^{3}}\int_{0}^{\infty}dE \frac{E^{3}}{e^{\beta E} - 1}\frac{1}{(1+\lambda E)^{4}}.
\eeq

This integral cannot be solved analytically. The approximation $\lambda k_{B} T \ll 1$ is introduced. This limit is justified for astrophysical objects since the highest temperatures attained in stars are around ($10^{11}$ $-$ $10^{12}$) K \cite{Padmanabhan:2001}, yielding the condition $\lambda\ll 10 $ GeV$^{-1}$ which is satisfied if $\lambda \sim E_{QG}^{-1}$ where $E_{QG}$ is the quantum gravity energy scale. The limit, $\lambda k_{B} T \gg 1$, presumably relevant at the early universe, was investigated in Ref. \cite{Alexander:2001ck}.

Defining a variable $y=\lambda E$. Eq. (\ref{eq:uintdef}) can be written as 

\beq\label{eq:uintdefy}
u=\frac{1}{\pi^{2}\hbar^{3}c^{3}\lambda^{4}}\int_{0}^{\infty}dy \frac{y^{3}}{e^{\frac{y}{\lambda k_{B}T}} - 1}\frac{1}{(1+y)^{4}}.
\eeq

As $\lambda k_{B}T\ll 1$, it suffices to expand $y^{3}/(1+y)^{4}$ in Taylor series around zero, since away from the origin other values will be exponentially suppressed by 
$(e^{\frac{y}{\lambda k_{B}T}} - 1)^{-1}$. One gets\footnote{Using the formula $
\int_{0}^{\infty}\frac{x^{\nu -1}\;dx}{e^{\mu x}-1}=\frac{1}{\mu^{\nu}}\Gamma{(\nu)}\zeta{(\nu)} \;\;\;[\mbox{Re}\mu > 0,\mbox{Re}\nu > 0],
$ where $\Gamma{(\nu)}$ and $\zeta{(\nu)}$ are gamma and zeta functions, respectively \cite{Gradshteyn:2000}.}

\beq\label{eq:uintdefyexp}
u=\frac{1}{\pi^{2}\hbar^{3}c^{3}\lambda^{4}}\left[\int_{0}^{\infty}dy \frac{y^{3}}{e^{\frac{y}{\lambda k_{B}T}} - 1}-4\int_{0}^{\infty}dy \frac{y^{4}}{e^{\frac{y}{\lambda k_{B}T}} - 1}\right]=\frac{4\sigma}{c}T^{4} - \frac{96\zeta(5)}{\pi^{2}\hbar^{3} c^{3}}\lambda k_{B}^{5}T^{5},
\eeq

\noindent where $\sigma=\frac{\pi^{2}k_{B}^{4}}{60\hbar^{3} c^{2}}$ is the Stefan-Boltzmann constant. Notice that the Stefan-Boltzmann law is recovered for $\lambda \rightarrow 0$. This expression can be rewritten as $u= 4\sigma_{\mbox{eff}}(T;\lambda)T^{4}/c$, where the effective Stefan-Boltzmann ``constant" is given by:

\beq
\sigma_{\mbox{eff}}(T;\lambda)=\sigma\left(1-\frac{1440\zeta(5)}{\pi^{4}}\lambda k_{B} T \right).
\eeq

So the leading noncommutative correction reduces the energy density. Using Eq. (\ref{eq:pressure}), the pressure is

\beq\label{eq:pressurerad}
P=\frac{1}{3\pi^{2}\hbar^{3}c^{3}}\int_{0}^{\infty}dE\;\frac{E^{3}}{e^{\beta E} -1}\frac{1}{(1+\lambda E)^{3}}.
\eeq

This integral can be solved in the same way as the energy density integral. One obtains

\begin{equation}\label{eq:pressurerada}
P=\frac{1}{\pi^{2}\hbar^{3}c^{3}\lambda^{4}}\left[\frac{1}{3}\int_{0}^{\infty}dy \frac{y^{3}}{e^{\frac{y}{\lambda k_{B}T}} - 1}-\int_{0}^{\infty}dy \frac{y^{4}}{e^{\frac{y}{\lambda k_{B}T}} - 1}\right]=\frac{4\sigma}{3c}T^{4}-\frac{24\zeta(5)}{\pi^{2}\hbar^{3}c^{3}}\lambda k_{B}^{5}T^{5}. 
\end{equation}

To find an Equation of State (EoS), one must divide Eq. (\ref{eq:pressurerada}) by Eq. (\ref{eq:uintdefyexp})

\beq\label{pdividedbyu}
\frac{P}{u}=\frac{1}{3}+(\lambda k_{B}T)\frac{120\zeta(5)}{\pi^{4}},
\eeq

Using Eq. (\ref{eq:uintdefyexp}), one writes $T$ as a power series in $\lambda$. Substituting this result into Eq. (\ref{pdividedbyu}) one finally gets the leading noncommutative correction to the  EoS 

\begin{equation}\label{eq:nceosrada}
P=\frac{u}{3}\left[1+\frac{360\zeta(5)}{\pi^{4}}\left(\frac{15}{\pi^{2}}\right)^{1/4}\lambda (\hbar c)^{3/4} u^{1/4}\right].
\end{equation}

\noindent Notice that the usual relationship $u=3P$ is recovered for $\lambda\rightarrow 0$.

\subsection{Non-relativistic Ideal Gas}

Consider Eq. (\ref{eq:folambda}), hence for a non-relativistic ideal gas $p\ll mc$, it reads

\beq
E=\lambda p^{2}c^{2} + mc^2 + \frac{p^{2}}{2m} + \mathcal{O}\left(\frac{p}{mc}\right)^{4}.
\eeq

The condition to be satisfied in order to ensure the relevance of the noncommutative correction is $\lambda\;\laq\;(2mc^{2})^{-1}$. However, for matter found in main-sequence stars, hydrogen for simplicity, one finds $\lambda\;\laq\; 1$ GeV$^{-1}$, which certainly is a too stringent bound. Hence, one does not consider any noncommutative correction to the non-relativistic ideal gas. The pressure is the well know expression \cite{Landau:1980} $P=Nk_{B}T/V$. After introducing the mean molecular weight $\mu_{N}=\rho/(n m_{N})$, where $n=N/V$, $\rho$ is the mass density and $m_{N}$ is the nucleon mass\footnote{$m_{N}=1 \mbox{amu} = 931,494\; \mbox{MeV/c}^{2}$\cite{Amsler:2008zzb}.}, the pressure reads

\beq\label{eq:nrigpressurefinal}
P=\frac{\rho}{\mu_{N} m_{N}}k_{B}T.
\eeq

\subsection{Deformed Degenerate Fermion Gas} \label{subsec:degfergas}
A fermion gas becomes degenerate at low temperatures ($T\rightarrow 0$). In this case, the fermionic character gives origin to a very simple expression for the occupation number. In terms of the momentum, the occupation number is $n(p)=H(p_{F}-p)$, where $p_{F}$ is the Fermi momentum, above which there are no occupied states and $H(x)$ is the Heaviside step function. For stars, the momentum varies between MeV/c $-$ GeV/c, so $p_{F}\ll (\lambda c)^{-1}$ and one uses Eq. (\ref{eq:folambda}) as the deformed dispersion relation.

For spin one-half particles ($\gamma=2$), the particle number density Eq. (\ref{eq:numberdensity})  is given by

\beq \label{eq:numberdensityfermion}
n=\frac{N}{V}=\frac{1}{\pi^{2}\hbar^{3}}\int_{0}^{\frac{1}{\lambda c}}dp n(p) p^{2} = \frac{1}{\pi^{2}\hbar^{3}}\int_{0}^{p_{F}}dp  p^{2} =\frac{p_{F}^{3}}{3\pi^{2}\hbar^{3}}=\frac{(mc)^{3}x^{3}}{3\pi^{2}\hbar^{3}},
\eeq

\noindent where $x=p_{F}/mc$. Using Eqs. (\ref{eq:folambda}) and (\ref{eq:energydensity}), one obtains the energy density

\begin{eqnarray}
u&=&\frac{E}{V}=\frac{1}{\pi^{2}\hbar^{3}}\int_{0}^{p_{F}}dp\;p^{2}\sqrt{p^{2}c^{2}+m^{2}c^{4}} + \frac{\lambda c^{2}}{\pi^{2}\hbar^{3}}\int_{0}^{p_{F}}dp\;p^{4} \nonumber \\
&=&\frac{(mc^{2})^{4}}{\pi^{2}(\hbar c)^{3}}\left(\frac{x^{3}\sqrt{1+x^{2}}}{3}-\frac{f(x)}{24} +(\lambda mc^{2})\frac{x^{5}}{5}\right).\label{eq:energydensityfermionc}
\end{eqnarray}

\noindent where $f(x)=x(2x^{2}-3)\sqrt{1+x^{2}}+3\sinh^{-1}x$ \cite{Chandrasekhar:1967}. This shows that the first noncommutative correction is given by $\lambda mc^{2}$. The pressure is obtained using Eqs. (\ref{eq:folambda}) and (\ref{eq:pressure}):

\beq\label{eq:pressurefermionc}
P=\frac{c^{2}}{3\pi^{2}\hbar^{3}}\int_{0}^{p_{F}}dp\;\frac{p^{4}}{\sqrt{p^{2}c^{2}+m^{2}c^{4}}} + \frac{2\lambda c^{2}}{3\pi^{2}\hbar^{3}}\int_{0}^{p_{F}}dp\;p^{4}=\frac{(mc^{2})^{4}}{\pi^{2}(\hbar c)^{3}}\left(\frac{f(x)}{24} +(\lambda mc^{2})\frac{2x^{5}}{15}\right).
\eeq


\section{Application to Astrophysical Objects}\label{sec:aao}

\subsection{Main-sequence Stars: The Sun}\label{sec:msssun}

The simplest model for a main-sequence star consists in assuming that the star is composed by a non-relativistic ideal gas and radiation that are maintained in equilibrium by gravity \cite{Chandrasekhar:1967}. The total pressure $P$ comprises the sum of the radiation pressure $P_{\mbox{rad}}$ (Eq. (\ref{eq:pressurerada})) and the gas pressure $P_{\mbox{gas}}$ (Eq. (\ref{eq:nrigpressurefinal})). One denotes the ratio between $P_{\mbox{gas}}$ and $P$ by $\beta_{S}$. One then has $P=P_{\mbox{gas}}/\beta_{S}=	P_{\mbox{rad}}/(1-\beta_{S})$. Substituting Eqs. (\ref{eq:pressurerada}) and (\ref{eq:nrigpressurefinal}) into this relationship one finds

\beq\label{eq:pressureradgas}
\frac{\rho}{\mu_{N} m_{N} \beta_{S}}k_{B}T=\frac{4\sigma}{3c(1-\beta_{S})}T^{4}-\lambda\frac{24\zeta(5)k_{B}^{5}}{\pi^{2}\hbar^{3}c^{3}(1-\beta_{S})}T^5,
\eeq

The aim is to obtain $P=P(\rho)$, so one must solve this equation for $T$, obtaining an expression $T=T(\rho)$ up to the first order in $\lambda$. Inserting this result into $P_{\mbox{gas}}=\beta P$ and using Eq. (\ref{eq:nrigpressurefinal}) one finds that $P=K_{1}\rho^{4/3}+\lambda K_{2}\rho^{5/3}$ where

\begin{eqnarray}
K_{1}&=&\frac{(1-\beta_{S})^{1/3}}{(\beta_{S}\mu_{N})^{4/3}}\left(\frac{3ck_{B}^{4}}{4\sigma m_{N}^{4}}\right)^{1/3}=2.67\times 10^{10}\left[\frac{(1-\beta_{S})^{1/3}}{(\beta_{S}\mu_{N})^{4/3}}\right] \;\frac{\mbox{J m}}{\mbox{kg}^{4/3}} \label{eq:k1}\\
K_{2}&=&\frac{(1-\beta_{S})^{2/3}}{(\beta_{S}\mu_{N})^{5/3}}\left(\frac{360\zeta(5)}{\pi^{4}}\right)\left(\frac{3ck_{B}^{4}}{4\sigma m_{N}^{5/2}}\right)^{2/3}=4.52\times 10^{-6}\left[\frac{(1-\beta_{S})^{2/3}}{(\beta_{S}\mu_{N})^{5/3}}\right] \;\frac{\mbox{J$^{2}$ m$^{2}$}}{\mbox{kg}^{5/3}}\label{eq:k2}.
\end{eqnarray}

One assumes the Eddington's Standard Model of Stars \cite{Padmanabhan:2001}: $\beta_{S}$ is constant inside the star, and $\mu_{N}$ is constant since the chemical composition is not supposed to change. Hence, one finds that the pressure is a contribution of two polytropes: a $n=3$ polytrope for the mixture of gas and radiation and a $n=3/2$ polytrope for the NC contribution, which is considered as a perturbation. The question of stability can be understood as follows. Defining $\Gamma=1+\frac{1}{n}$, a $n=3$ ($\Gamma=4/3$) polytrope represents a marginally stable configuration \cite{Padmanabhan:2001}. If one introduces corrections due to general relativity, one shows that, in fact, this polytrope is unstable. One follows here the purely Newtonian analysis \cite{Padmanabhan:2001}. The total energy is written as the sum of the internal energy ($\sim PV$) and the gravitational potential energy ($\sim \frac{GM^{2}}{R}$). The pressure is given by $P=K_{1}\rho^{\Gamma_{1}}+\lambda K_{2}\rho^{\Gamma_{2}}$ and to ensure that the second term is a perturbation $\lambda\frac{K_{2}}{K_{1}}\rho^{\Gamma_{2}-\Gamma_{1}}\ll 1$. The energy can be written as

\beq
E =k_{0}PV-k_{1}\frac{GM^{2}}{R} =C_{1}M\rho_{c}^{\Gamma_{1}-1}+C_{2}M\rho_{c}^{\Gamma_{2}-1}-k_{3}M^{5/3}\rho_{c}^{1/3},
\eeq

\noindent where $k_{0},k_{1},k_{3},C_{1},C_{2}$ are constants and $\rho_{c}$ is the central density. The value of $M$ at the critical point $\left( \frac{\partial E}{\partial \rho_{c}}\right)=0$ is computed and one performs

\beq
\frac{d\ln M}{d\ln \rho_{c}}=\frac{3}{2}\left(\Gamma_{1}-\frac{4}{3}\right)+\lambda\frac{3C_{2}(\Gamma_{2}-1)}{2C_{1}(\Gamma_{1}-1)}(\Gamma_{2}-\Gamma_{1})\rho_{c}^{\Gamma_{2}-\Gamma_{1}}.
\eeq

A necessary, but not sufficient, condition for stability of the configuration is that $\frac{d\ln M}{d\ln \rho_{c}}>0$, for $M$ given by $\frac{\partial E}{\partial \rho_{c}}=0$ \cite{Glendenning:1997wn}. Here, $\Gamma_{1}=4/3$ and $\Gamma_{2}=5/3>\Gamma_{1}$, so $\frac{d\ln M}{d\ln \rho_{c}}>0$. Thus, the effect of noncommutativity is to turn this configuration into a stable one.

To quantify the relevance of noncommutativity for these type of stars one chooses the Sun, as it is a very typical main-sequence star. In the model investigated here,  at the center, $1-\beta_{S} \approx 10^{-3}$, $\rho_{c}=1.53\times10^{5}$ kg/m$^{3}$ and $\mu_{N}=0.829$ \cite{Padmanabhan:2001}. The ratio between the noncommutative and the commutative contributions to the pressure is

\beq \label{eq:ratiopncpcmss}
\frac{P_{NC}}{P_{C}}=\frac{\lambda K_{2}\rho^{5/3}}{K_{1}\rho^{4/3}}=  9.66\times10^{-16}\lambda,
\eeq

\noindent where $P_{NC}=\lambda K_{2}\rho^{5/3}$ is the noncommutative correction to the pressure, $P_{C}=K_{1}\rho^{4/3}$ is the usual (commutative) term and $\lambda$ must be expressed in S.I. units. One computed this value at the center of the star using Eqs. (\ref{eq:k1}) and (\ref{eq:k2}). The value of $\lambda$ will be discussed later (Section \ref{sec:discussions}), but one sees that this ratio is fairly small.

\subsection{White Dwarfs}

As one has seen, for main-sequence stars the noncommutative correction is rather small and, even though it plays a role in the stability of the star,  one should seek more energetic configurations. White dwarfs are stars that have masses comparable to th Sun and planet sizes \cite{Bhatia:2001}. At their center, the range of possible mass densities is $10^{8}$ kg/m$^{3}$ $\laq\; \rho_{c}\;\laq\;10^{12}$ kg/m$^{3}$. At these densities the matter is completely ionized and the electrons are free. The treatment of matter must necessarily involve quantum mechanics. The simplest model of white dwarfs consists in considering an ideal electronic gas maintained in equilibrium by gravity. Although the central temperature is $T\sim10^{7}$ K, it is much smaller than the temperature associated with the Fermi energy ($T\sim10^{10}$ K). This justifies describing white dwarfs, at first approximation, as a degenerate electron gas.

Defining the electronic molecular weight as $\mu_{e}= \rho/n_{e}m_{p}$, where $n_{e}$ is the electronic particle number density and $m_{p}$ is the proton mass. One can rewrite Eq. (\ref{eq:numberdensityfermion}) as

\beq\label{eq:xwd}
x=(3\pi^{2})^{1/3}\frac{\hbar}{m_{e}c}n_{e}^{1/3}=\left(\frac{3\pi^{2}}{m_{p}}\right)^{1/3}\frac{\hbar}{m_{e}c}\left(\frac{\rho}{\mu_{e}}\right)^{1/3}=10^{-3}\left(\frac{\rho}{\mu_{e}}\right)^{1/3},
\eeq

\noindent where $m_{e}$ is the mass of the electron and this result is presented in S.I. units. One considers $\mu_{e}=2$, as this corresponds to a star composed mainly by Helium. For the possible values of the central density, one chooses a low density value $\rho=10^{8}$ kg/m$^{3}$, an intermediate density ($p_{F}\sim m_{e}c$) value $\rho=2\times10^{9}$ kg/m$^{3}$ and a high one $\rho= 10^{12}$ kg/m$^{3}$. One finds $x_{8}=0.37$, $x_{9}=1.01$, $x_{12} = 7.99$, respectively. One now quantifies the relevance of noncommutativity, using Eq. (\ref{eq:pressurefermionc}):

\beq\label{eq:pertpressurewd}
\frac{P_{NC}}{P_{C}}=\frac{48x^{5}}{15 f(x)} \lambda m_{e}c^{2}.
\eeq

\begin{figure}
\centering
\includegraphics[width=7cm]{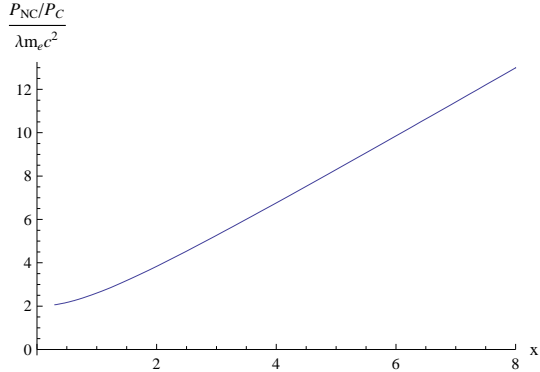}
\caption{Ratio between the noncommutative  and the commutative contributions to the pressure in units of $\lambda m_{e}c^{2}$ for white dwarfs and neutron stars.}
\label{fig:fermi}
\end{figure}

This function is plotted in Fig. \ref{fig:fermi} for the usual values of $x$ found in white dwarfs. One sees that the relevance of noncommutativity is just a few $\lambda m_{e}c^{2}$. Since $m_{e}c^{2}\sim 0.5$ MeV and $\lambda \sim (E_{QG})^{-1}$ (see Section \ref{sec:discussions}), this correction is much smaller than other improvements to the simplest white dwarf model, such as Coulomb interaction for low densities, and general relativity for high densities. Also  neglected are the corrections due to rotation and magnetic fields, which are likely to be greater than the noncommutative correction.

Let us now analyse the problem of stability (for details see e.g. Ref. \cite{Padmanabhan:2001}): the noncommutative correction is proportional to $x^{5} \propto \rho^{5/3}$ (Eq. (\ref{eq:xwd})) which represents a $\Gamma=5/3$ polytrope. This lies in the region of stability and hence noncommutativity does not bring any instability for white dwarfs.

\subsection{Neutron Stars}

Consider now that $\rho \;\gaq\; 10^{12}$ kg/m$^{3}$, a density for which the protons and electrons combine and through the weak interaction form neutrons. These objects must be described by general relativity since $\frac{GM}{Rc^{2}}\sim0.1$. Their masses are about the solar mass ($M\sim1 M_{\odot}$) and their radius about $R\sim 10$ km.

These stars are mainly composed by neutrons, hence the simplest model for a neutron star, the Oppenheimer-Volkoff (OV) model \cite{Oppenheimer:1939ne}, considers an ideal neutron gas counterbalanced by gravity. This model is rather unrealistic as one must introduce nuclear interaction effects. However, this simpler formulation has some advantages, the first being that one can compute the thermodynamical quantities in a closed and analytical form. Furthermore, the obtained results  do not differ significantly from a more realistic EoS \cite{Glendenning:1997wn}; finally the OV EoS is stiffer than other EoS that include nuclear interaction \cite{Fraga:2001xc}. This neutron gas is degenerate since the Fermi energy of neutrons ($\sim 1$ GeV) is much greater than the thermal energy associated to the gas ($\sim 10^{-1}$ MeV).

Using Eqs. (\ref{eq:numberdensityfermion}), (\ref{eq:energydensityfermionc}) and (\ref{eq:pressurefermionc}), one obtains

\begin{eqnarray}
x&=&\frac{\hbar}{m_{n}c}\left(3\pi^{2} n\right)^{1/3}, \label{eq:xneutron} \\
u&=&\frac{(m_{n}c^{2})^{4}}{\pi^{2}(\hbar c)^{3}}\left(\frac{x^{3}\sqrt{1+x^{2}}}{3}-\frac{f(x)}{24} +(\lambda m_{n}c^{2})\frac{x^{5}}{5}\right),\label{eq:energydensityneutron} \\
P&=&\frac{(m_{n}c^{2})^{4}}{\pi^{2}(\hbar c)^{3}}\left(\frac{f(x)}{24} +(\lambda m_{n}c^{2})\frac{2x^{5}}{15}\right),\label{eq:pressureneutron}
\end{eqnarray}

\noindent where $m_{n}$ is the neutron mass. One cannot obtain the EoS $P=P(u)$ in an analytical form. The relevance of noncommutativity is depicted in Fig. \ref{fig:fermi}, with the modification $m_{e}\rightarrow m_{n}$. For the usual baryonic matter, $n=0.15\times 10^{45}$ m$^{-3}$ \cite{Glendenning:1997wn}, one obtains $x=0.35$ and inserting this value into Eq. (\ref{eq:pressureneutron}), one obtains that $\frac{P_{NC}}{P_{C}}=2.1(\lambda m_{n} c^{2})$. Notice that as $m_{n} \sim 10^{3}m_{e}$, the effect of noncommutativity is much more important for this class of stars than for white dwarfs.

For neutron stars the question of stability is more complex since general relativity is required. It involves solving numerically the Oppenheimer-Volkoff equation  with the OV EoS.


\section{Discussions and conclusions}\label{sec:discussions}

In this contribution, the most relevant effects of noncommutativity for astrophysical objects were estimated. Noncommutativity is introduced through a deformed dispersion relation and the relevant thermodynamical quantities were calculated using the grand canonical ensemble. These results are applied to physical models describing main-sequence stars, white-dwarfs and neutron stars.

In what concerns the noncommutative parameter, $\lambda$, (cf. Eqs. (\ref{eq:1})  and (\ref{eq:folambda})), an upper bound for its value can be inferred from the breaking of Lorentz invariance applied to the problem of ultra-high energetic cosmic rays. The most stringent limit is $\lambda<2.5\times 10^{-19}$ GeV$^{-1}$ $=1.6\times10^{-9}$ J$^{-1}$. For the stars, one is certainly in the low-energy limit, and hence this justifies the expansion of the quantities up to first order in $\lambda$. 

Using $\lambda\sim 10^{-19}$ GeV$^{-1}$, one can estimate the ratio between $P_{NC}$ and $P_{C}$ for the main-sequence stars (as an example one chooses the Sun), white-dwarfs and neutron stars. One obtains $\frac{P_{NC}}{P_{C}}\sim 10^{-25}$, $\frac{P_{NC}}{P_{C}}\sim 10^{-22}$ and $\frac{P_{NC}}{P_{C}}\sim 10^{-19}$ (cf. Eqs. (\ref{eq:ratiopncpcmss}), (\ref{eq:pertpressurewd})  and (\ref{eq:pressureneutron})), respectively. These results indicate that noncommutative correction is fairly small, actually smaller than other usually neglected standard effects in stellar physics. As one expects the relevance of noncommutativity to grow as one considers denser configurations, one is lead to conclude that noncommutative might be relevant for black holes. In fact, full phase-space noncommutativity is shown to be quite important for Schwarzschild black holes \cite{Bastos:2009ae}.

For main sequence stars and white-dwarfs, the question of stability has been analysed and shown that noncommutativity does not introduce any instability. Actually, for main sequence stars, noncommutativity is beneficial for the star stability.



\ack{The work of C. A. D. Z. is fully supported by the FCT fellowship SFRH/BD/29446/2006.}


\end{document}